\documentclass[showpacs,showemail,preprintnumbers,pre]{revtex4}
\usepackage{graphicx}
\usepackage{dcolumn}
\usepackage{bm}
\newcommand{\y }{\'{\i}}
\begin{document}

\begin{center}
{\bf RESPONSE OF 
A MODEL OF CO OXIDATION WITH CO DESORPTION AND DIFFUSION TO A 
PERIODIC EXTERNAL CO PRESSURE}\\
\vspace{15pt}
G.~M.Buend\y a and E.~Machado\\
{\it Physics Department, Universidad Sim\'on Bol\y var,\\
Apartado 89000, Caracas 1080, Venezuela}\\
\vspace{5pt}
P.~A.~Rikvold\\
{\it Center for Materials Research and Technology,\\
School of Computational Science,  and Department of Physics, \\
Florida State University, Tallahassee, Florida 32306-4120, USA}\\
\end{center}

\section*{Abstract}

We present a study of the dynamical behavior of a
Ziff-Gulari-Barshad model with CO desorption and lateral diffusion.
Depending on the values of the desorption and diffusion parameters, 
the system presents a discontinuous phase transition between low and high CO coverage phases. 
We calculate several points on the coexistence curve between these phases. 
Inclusion of the diffusion term produces a significant increase in the CO$_2$ production rate.
We further applied a square-wave periodic pressure
variation of the partial CO pressure with parameters that can be tuned to modify the catalytic activity. 
Contrary to the diffusion-free case, this driven system does not present a 
further enhancement of the catalytic activity, beyond the increase induced by the diffusion under constant 
CO pressure.

\section{Introduction}

Phase transitions and critical phenomena in open nonequilibrium 
statistical systems have attracted a great deal of interest in recent decades, due to their
wide applications in many branches of physics, chemistry, biology, economics, and 
even sociology \cite{JENS98}. Among these, surface reaction models have become
the archetype to study out-of-equilibrium critical phenomena, and they have been intensely
analyzed with the purpose of designing more efficient catalytic processes \cite{CHRI94}. The Ziff,
Gulari, Barshad (ZGB) model \cite{ZIFF86} with desorption (ZGB-k) 
\cite{TOME93,EHSA89,MACH05A,MACH05B} describes some kinetic aspects of the
reaction CO+O$_2$ $\rightarrow$ CO$_2$ on
a catalytic surface in terms of two parameters: the partial
pressure of CO, $y$, which represents the probability per unit of time that the next
molecule arriving at the surface is a CO, and the desorption parameter, $k$, which gives 
the probability per unit of time that a site occupied by a CO molecule is vacated.  
Below a critical CO desorption rate, the model
exhibits a first-order phase transition between low and high coverage phases 
\cite{TOME93,ZIFF92,MACH05A,MACH05B}. 
A study of the metastable lifetimes associated with the transition between 
these phases indicates that the decay mechanism is very similar to the decay
of metastable phases associated with equilibrium first-order phase transitions  
and can be described by the classic Kolmorogov-Johnson-Mehl-Avrami (KJMA) theory \cite{KJMA} of 
phase transformations by nucleation and growth \cite{MACH05B}.
Near the coexistence line between the active and the CO poisoned phase, the decay times of the metastable state 
are different if the ZGB-k model is driven into the active phase from 
the CO poisoned state or vice versa \cite{MACH05B}.
This asymmetry of the lifetimes, together with studies that indicate that  
it is possible to increase the efficiency of catalytic reactions 
by forcing the system by a periodic external pressure that switches back and
forth around the discontinuous transition \cite{IMBI95,LOPE00,HUA02}, 
inspired us to subject the system to a square-wave periodic CO pressure that stays for a time 
$t_d$ in the high-production region and $t_p$ in the low-production region. We showed that 
$t_p$ and $t_d$ can be tuned to significantly enhance the time-averaged catalytic activity of the system \cite{MACH05A}.

Many experiments indicate that at intermediate and high temperatures adsorbed CO is fairly mobile, 
while oxygen is still relatively immobile ~\cite{EHSA89,IMBI95}. In this work we aim at a more realistic 
process by incorporating CO diffusion into the ZGB-k model. The model that includes CO desorption and 
diffusion will be denoted the ZGB-(k,d) model. We investigate how the diffusion process 
affects the first-order kinetic phase transition and the response of the system to a periodic variation 
of the CO pressure.

The rest of this paper is organized as follows: in Sec.~II we define the
model and describe the Monte Carlo simulation techniques used. In
Sec.~III we present and discuss the results obtained when
subjecting the model to a periodic external CO pressure. Finally,
we present our conclusions in Sec.~IV.

\section{Model and simulations}

The ZGB-(k,d) model is simulated on a square lattice of linear size $L$ that represents the catalyst surface. 
A Monte Carlo simulation generates a sequence of trials: CO diffusion with probability per unit of time $d$, and 
adsorption or desorption with $(1-d)$. In nature, the diffusion process occurs more frequently than 
the adsorption/desorption process, so that $d \in (1/2,1)$ \cite{PAVL01}. 
The chemical processes include  CO or O$_2$ adsorption with probability per unit of time $1-k$, and CO desorption with 
probability per unit of time $k$. For the adsorption, a CO or O$_2$ molecule is selected with probability per unit of time $y$ and $1-y$ respectively. 
The algorithm works in the following way. A site $i$ is selected at random. In the case of CO diffusion, 
if $i$ is empty or occupied by an O atom, the trial ends. Otherwise, a nearest neighbor of $i$ is selected at random; 
if it is vacant, the CO molecule exchanges with it, and reacts if any of its nearest-neighbors is an O atom, 
producing a CO$_2$ molecule and leaving
two empty sites on the surface. In the case of desorption, if $i$ is occupied by a CO, the site $i$
is vacated and the trial ends; if not, the trial also ends. In the case of adsorption, if
a CO molecule is selected it can be adsorbed at the empty site $i$ if none of its nearest neighbors are 
occupied by an O atom. Otherwise, one of the occupied O neighbors is selected
at random and removed from the surface, leaving $i$ and the selected neighbor vacant. O$_2$
molecules require a nearest-neighbor pair of vacant sites to
adsorb. Once the O$_2$ molecule is adsorbed, it is dissociated into
two O atoms. If an O atom is located next to a site filled with a
CO molecule, they react to form a CO$_2$ molecule that escapes,
leaving two sites vacant. This process simulates the CO $+$ O $\rightarrow$ CO$_2$ surface reaction. 
Since the diffusion is much faster than the other chemical processes, we will assume that the diffusion does 
not slow down the time scale for adsorption/desorption, which is the physically relevant time scale for the phase 
transformation \cite{FRAN05}. Thus, we define the time unit as one Monte Carlo Step per Site, MCSS, in which 
the system performs, on average, one adsorption/desorption attempt per site.

\section{Results}

\begin{figure}[ht]
\includegraphics[angle=0, width=.45\textwidth]{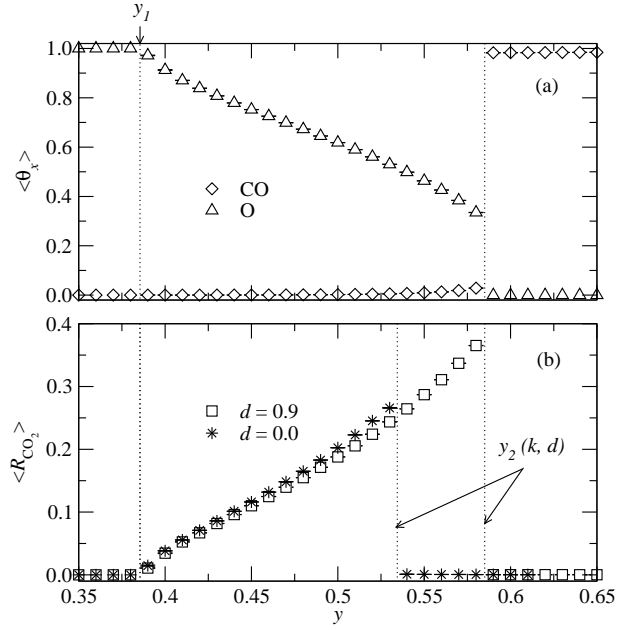}\\
\caption[]{(a) Average values of the CO and O coverages, $\theta_{\rm CO}$
and $\theta_{\rm O}$, shown as functions of the stationary applied CO pressure
$y$, for $L=100$ with $k=0.01$ and $d=0.9$. A continuous, nonequilibrium
phase transition occurs at $y_1$, and a discontinuous one at $y_2(k,d)$. 
(b) Comparison between the average 
CO$_2$ production rate $\langle R_{\rm CO_2}\rangle$ at $d=0.9$ and $d=0$, calculated 
for the same values of $L$ and $k$ as in (a). Both $y_2$ and the maximum value of $\langle R_{\rm CO_2}\rangle$ 
increase significantly with $d$.}
\label{fig1}
\end{figure}

\begin{figure}[ht]
\vspace{1.0truecm}
\includegraphics[angle=0, width=.45\textwidth]{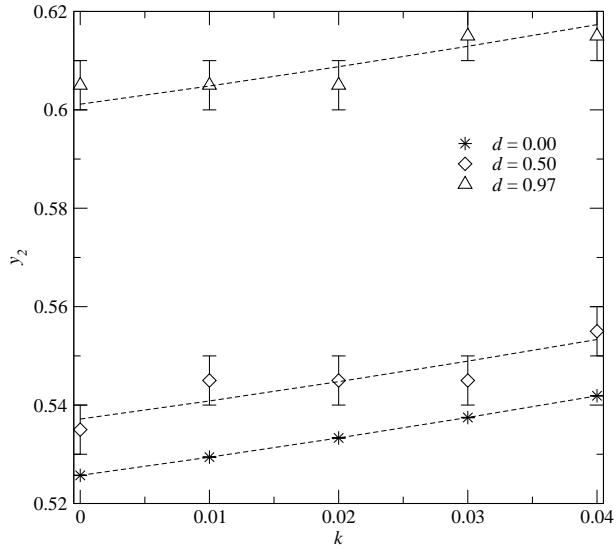}\\
\caption{Some points on the coexistence curve for several values of the diffusion probability per unit of time $d$. 
This is analogous to the pressure vs temperature phase diagram for a fluid in equilibrium. 
The continuous dashed lines represent a quadratic fit to the data.}
\label{fig2}
\end{figure}

Our simulations are realized on a square lattice of
$100{\times}100$ sites, assuming periodic boundary conditions.
We calculate the  coverages $\theta_{\rm CO}$ and
$\theta_{\rm O}$,  defined as the fraction of surface sites occupied by CO and O,
respectively, and $R_{{\rm CO}_2}$ defined as the rate of production of CO$_{2}$.
In Fig.~\ref{fig1} we show, for a particular value of $k$ and $d$, the
dependence of the average values of the coverages and the
production rate: $\langle \theta_{\text{CO}}\rangle$,
$\langle\theta_{\text{O}}\rangle$, and $\langle R_{\text{CO}_2}\rangle$, respectively,
for an external constant CO pressure $y$. There are two inactive 
regions, $y<y_1$ ($y_1$ appears to be fairly independent of $k$ and $d$ 
\cite{MACH05B}), and
$y>y_2(k,d)$. These correspond to the cases in which the surface is
saturated with O and CO, respectively. In between, for
$y_{1}<y<y_{2}(k,d)$, there is an active window where the system
produces CO$_{2}$. Notice that the maximum value of
$\langle R_{\text{CO}_2}\rangle$ is reached as $y_2(k,d)$ is approached from the
low-$\theta_{\text{CO}}$ phase. Fig.~\ref{fig1}(b) shows that the effect of the CO diffusion is to widen the 
range of values of $y$ for which there is catalytic 
activity, producing a significant increase of the maximum CO$_2$ production rate. 
In Fig.~\ref{fig2} we plot the coexistence point, $y_2$, as a function of $k$ for several values of $d$. 
The effect of the diffusion is to shift the transition to higher values of $y_2$, 
while maintaining approximately the same dependence of $y_2$ on $k$ that is observed at $d=0$.

\begin{figure}[ht]
\includegraphics[angle=0, width=.45\textwidth]{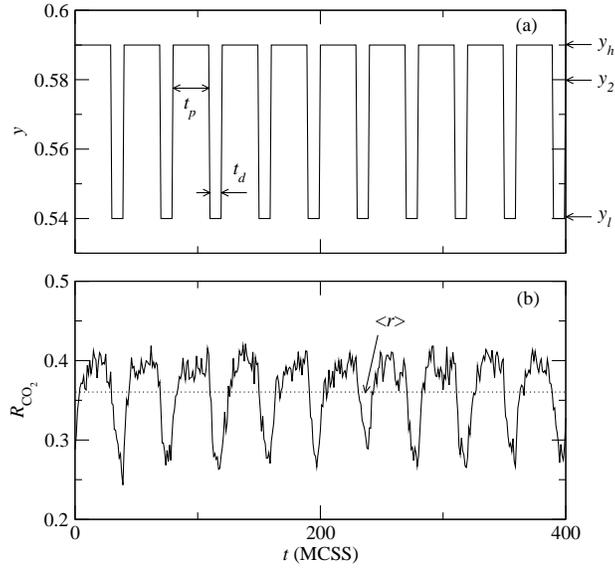}\\
\caption{(a) Applied periodic pressure of CO, $y(t)$,
which takes the values
$y_l=0.54$ and $y_h=0.59$ during the time intervals $t_d=10$ and
$t_p=30$, respectively.
(b) Response of the CO$_2$ production rate to the
applied pressure given in (a) for $k=0.01$ and $d=0.9$.
The dotted line marked $\langle r \rangle $
indicates the long-time average of the
period-averaged CO$_2$ production
rate $r$. In contrast to the case for $d=0$, this value cannot be distinguished from
the maximum average CO$_2$ production rate for {\it constant\/} $y=y_2(k,d)$, shown
in Fig.~\ref{fig1}.
Time is measured in units of Monte Carlo steps per site.}
\label{fig3}
\end{figure}

Next, we perturb the system by applying an oscillating pressure $y(t)$
that in a period $T=t_{d}+t_{p}$ takes the values ,
\begin{equation}
y=\left\{
\begin{array}{ll}
y_l & \mbox{during the time interval}\ t_d \\
y_h & \mbox{during the time interval}\ t_p
\;,
\end{array}
\right.
\label{y_vs_t}
\end{equation}
located at both sides of the transition point, i.e., $y_l < y_2(k)
<y_h$ (see Fig.~\ref{fig3}(a)).

We found that, for each choice of $y_l$ and $y_h$, we can vary the
productivity of the system by tuning
$t_{d}$ and $t_{p}$, the times that the driving force spends in
the high and low coverage regions, respectively. 
In Fig.~\ref{fig3}(b) it can
be seen that the response to the periodic pressure
shown in Fig.~\ref{fig3}(a), the CO$_2$ production rate $R_{\text{CO}_2}$, also
exhibits an oscillatory behavior.  We define the period-averaged value of
$R_{\text{CO}_2}$,
\begin{equation}
r = \frac{1}{T}\int R_{\text{CO}_2}(t)dt,
\label{order_parameter}
\end{equation}
as the dynamic order parameter. For the parameters used in Fig.~\ref{fig3} (selected after an exhaustive search 
looking for the maximum productivity)
the long-time average of $r$, $\langle r \rangle$ is almost the same as the maximum average
CO$_2$ production rate for constant $y$. In contrast, the ZGB-k model
shows a marked enhancement of the catalytic activity when subjected to this type
of periodic perturbation \cite{MACH05A}. We think that this significant difference may possibly be 
due to the fact that the 
diffusion term modifies the decay mechanisms of the system near the transition. Previous studies in a 
lattice-gas model near an equilibrium first-order phase transition 
show that at high diffusion rates the nucleation no longer can be described by the 
classical KJMA theory  \cite{FRAN05}. The fact that applying a periodic CO pressure to the system with diffusion, does not seem to increase the CO$_2$ production further, may indicate that the optimal mixing of CO and O on the surface was already reached by the inclusion of diffusion.

\begin{figure}[ht]
\includegraphics[angle=0, width=.45\textwidth]{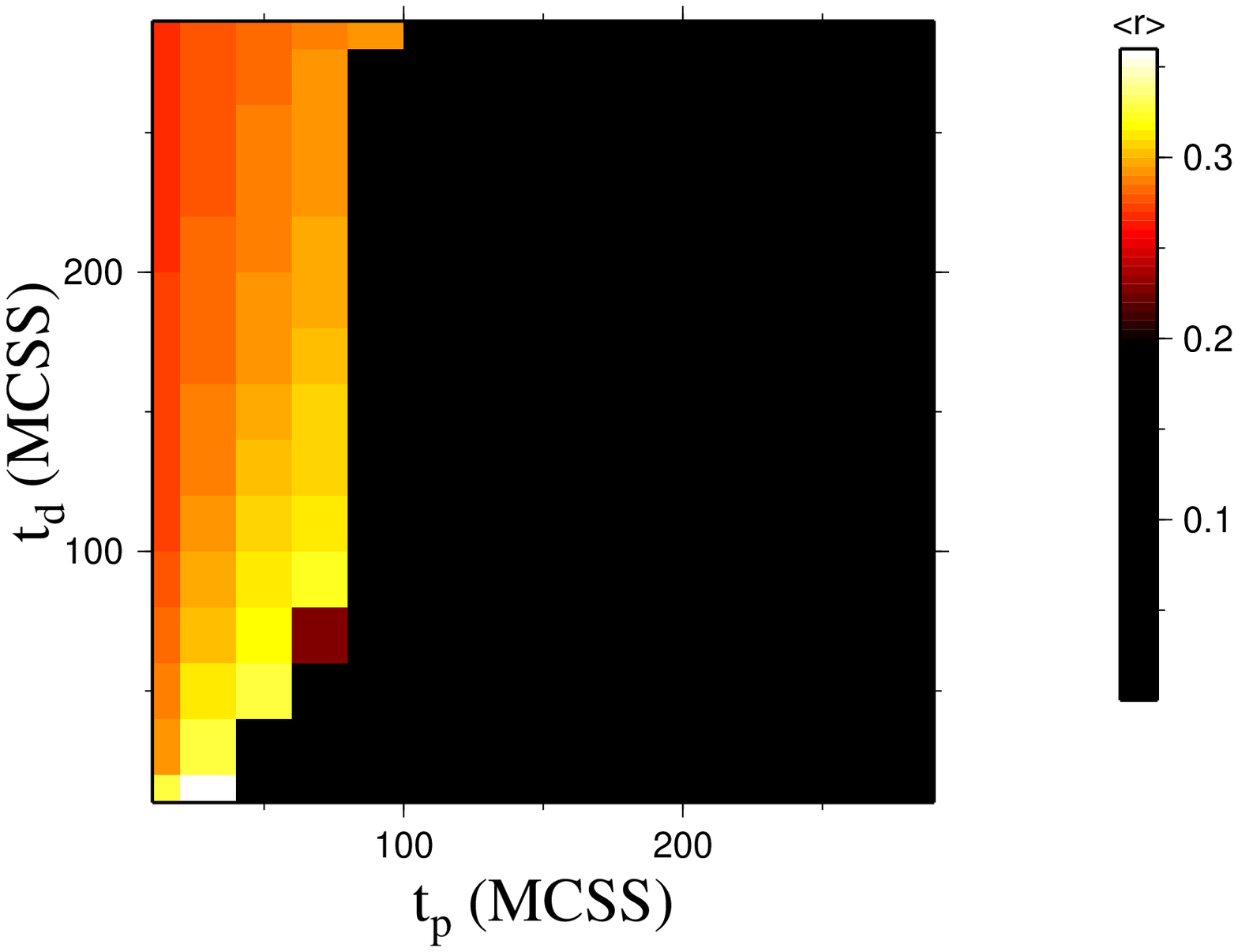}\\
\caption{Long-time average $\langle r\rangle$ of the period-averaged CO$_2$ production 
rate, shown as a density plot vs $t_d$ and $t_p$ for 
$y_l=0.54$, $y_h=0.59$, $k=0.01$, $d=0.9$ and $L=100$.
}
\label{fig4}
\end{figure}

Depending on the values of $t_d$ and $t_p$, the system behavior defines 
two regions: a productive one with
$r\ne 0$, and a non-productive one with $r\approx 0$, separated by a
coexistence curve. In Fig.~\ref{fig4}
we show a density plot of the average productivity $\langle r \rangle$ in
terms of $t_d$ and $t_p$ for a particular choice of $y_l$ and $y_h$.
Notice that the highest productivity is reached close to the
coexistence curve. The plot suggests, as is also the case in the ZGB-k model \cite{MACH05A},
the existence of a dynamic
phase transition (DPT), where the period-averaged order parameter $r$ takes the
value $r\approx 0$ or $r\ne 0$ depending on whether the system is in a
nonproductive state or in a productive one.
A more detailed study of the nature of the transition, based on 
a finite-size scaling analysis of the fluctuations of the order parameter, is
in progress and will be published elsewhere.

\section{Conclusions}
In this paper we have studied the ZGB model with
CO desorption and diffusion, the ZGB-(k,d) model, and its dynamic response to periodic variations of the partial 
CO pressure. We found that, like the model without diffusion (ZGB-k), this system exhibits a non-equilibrium 
discontinuous phase transition between low and high CO coverage phases. The transition occurs at a $k$-dependent 
value of the CO pressure $y_2$. The effect of the diffusion term is to shift the transition to a higher  
pressure, such that $y_2(k,d)>y_2(k,0)$, and to produce a significant increase of the maximum CO$_2$ production rate. 
We further subjected the system to a square-wave periodic CO pressure that switches back and forth around the
coexistence value that separates the low and high CO-coverage phases, staying 
for a time $t_d$ in the high-production region and for a time $t_p$ in the low production one. We found that 
$t_p$ and $t_d$ can be tuned to vary the catalytic activity of the system. However, contrary to the case of the 
ZGB-k model, the increase in the production of CO$_2$ obtained in this way is negligible. We believe that this result
indicates that the decay mechanisms of the metastable phase may be strongly affected by the diffusion process.
We also found indications that this driven nonequilibrium system could have a dynamic phase
transition between a dynamic phase of high CO$_2$ production and a
nonproductive phase. A more detailed study that includes the application of finite-size scaling techniques is 
necessary to establish the full details of the transition. 

\section*{Acknowledgments}
This work was supported in part by U.S.\ National Science 
Foundation Grant No.\ DMR-0240078 at Florida State University.


\end{document}